# A Real-time Robust Low-Frequency Oscillation Detection and Analysis (LFODA) System with Innovative Ensemble Filtering

Desong Bian, *Member*, *IEEE*, Zhe Yu, *Member, IEEE*, Di Shi, *Senior Member, IEEE*, Ruisheng Diao, *Senior Member*, *IEEE,* Zhiwei Wang, *Senior Member, IEEE*

*Abstract*—Low-frequency oscillations are hazardous to power system operation, which can lead to cascading failures if not detected and mitigated in a timely manner. This paper presents a robust and automated real-time monitoring system for detecting grid oscillations and analyzing their mode shapes using PMU measurements. A novel Extended Kalman Filtering (EKF) based approach is introduced to detect and analyze oscillations. To further improve the accuracy and efficiency of the presented software system, it takes advantages of three effective signal processing methods (including Prony's Method, Hankel Total Least Square (HTLS) Method, EKF) and adopts a novel voting schema to significantly reduce the computation cost. Results from these methods are processed through a time-series filter to ensure the consistency of detected oscillations and reduce the number of false alarms. The Density-Based Spatial Clustering of Applications with Noise (DBSCAN) method is used to accurately classify oscillation modes and the PMU measurement channels. The LFODA system has been functioning well in the State Grid Jiangsu Electric Power Company with 176 PMUs and 1000+ channels since Feb. 2018, demonstrating outstanding performance in reducing false alarms with much less computational cost.

*Index Terms*—Low frequency oscillation, PMU, Prony, HTLS, Extended Kalman Filter.

## I. Introduction

As power grid evolves for more than a century, today's power industry is facing new challenges due to the increasing penetration of renewable energy, distributed energy resources, new power market rules and behavior, demand response, electric vehicles, storage devices, power electronic devices and others. The modern grid behavior becomes more dynamic and stochastic, which were not observed before and now may invalidate many study assumptions typically used for grid planning and operation. For an extended period of time, large interconnected power network with big power transfers among balancing authorities or control areas would co-exist with micro-grids with various types of smart grid technologies. Of the many known challenges, grid oscillation remains a critical one that deserves much attention [1], especially when a large amount of power is transferred over a long geographical distance in a relatively weak power grid or at stressed operating conditions. Under such circumstances, higher risk of low-frequency oscillations can be observed. According to [2], low-frequency oscillations can be classified into local and global (inter-area) modes. Once excited, both types of oscillations become harmful when a power grid has insufficient ability to quickly dampen them out. A long period of local oscillations can destroy power equipment such as rotor shafts of generator. Moreover, inter-area oscillation with negative damping factors can cause cascading outages affecting a wide area in the grid [3]. A typical example was the WSCC system breakup event that occurred on August 10, 1996, in the Western Interconnection of North America, leaving ~7.5 million customers in the dark [4]. More recently, low-frequency oscillations can be observed in the tie lines connecting Mexico and Guatemala in Central America where grid operators have to frequently disconnect the transmission lines between the two countries in case of growing oscillations to avoid equipment damage and other severe issues.

Thus, a robust, accurate and fast low-frequency oscillation detection and analysis system running in real time is of critical importance to power system operators. Ideally, it should have the following desired function modules and key features: (a) accurately and quickly detecting oscillations of interest and providing in-depth diagnosis report including oscillation types, frequency, mode shape, damping factor, and location at their early stages, e.g., using ambient PMU data; (b) minimizing false alarms when the system is under normal operating conditions with effective noise processing and rejection; (c) having the capability of processing massive PMU channels for a large power grid in real time with minimal latency, which requires highly efficient computational algorithms and operation logic.

To achieve the above goals, researchers have developed various algorithms. One of the well-known algorithms is Prony that equips with excellent performance on event data with obvious oscillation modes; however, its performance may significantly degrade when applied to signals with a high noise level. It is also reported that the Matrix Pencil (MP) algorithm with singular value decomposition (SVD) was used to extract dynamic information from measurement data with noise [5] and detect low-frequency oscillations [6]. However, the high-level

This work has been accepted by CSEE Journal of Power and Energy Systems in 2019.

This work is supported by SGCC Science and Technology Program under project AI based oscillation locating.

D. Bian, Z. Yu, D. Shi, R. Diao, and Z. Wang are with GEIRI North America, San Jose, CA 95134. Email: di.shi@geirina.net.

noise in the signal will reduce the performance of the MP method [7]. In addition, the Hankel Total Least Square (HTLS) method was proposed, which can provide reasonable results at high signal-to-noise ratio but its performance degrades when the level of noise increases [8]. It is observed that most of the above methods are sensitive to noise. Nonetheless, the level of noise in PMU data is relatively high in real world. Therefore, this paper presents the EKF based method that has very good performance at low signal-to-noise ratio, though its accuracy and convergence rely heavily on the choice of initial points.

It is easy to observe that all these methods have a good potential for detecting oscillations and perform analysis when measurement data contains sufficient event information, e.g., obvious oscillation at various frequencies; however, their implementation can be limited in real world due to their high sensitivity level to measurement noise, leading to a large number of false alarms [9]-[11].

Consequently, oscillation detection and analysis implementing a single method alone is especially challenging. To overcome this issue, various platforms and tools for oscillation detection were developed. In [12], a post-event analysis tool for low-frequency oscillations was presented, which mainly targeted at offline applications. An online oscillation detection software was developed by ABB that focused on detecting inter-area oscillations only [13]. Recently, the Washington State University research team led by Prof. Mani Venkatasubramanian has developed a software platform for online oscillation detection and analysis [14]-[15]. It contains three different oscillation detection algorithms and the final decision for each time step is based on crosschecking results from all three methods. The system has a high level of accuracy but also comes with expensive computational cost, since data from each time period needs to be processed by all methods for further analysis. With the growing trend of PMU installation, e.g., in North America, processing all PMU channels in real time for a large power grid with massive data points can be a bottleneck affecting the online performance, especially during emergency conditions.

Inspired by the ideas in [14] and [15], a robust and fast Low-Frequency Oscillation Detection and Analysis (LFODA) software system with novel ensemble filtering is presented in this paper to further improve computational speed and robustness of the existing oscillation detection methods, for real-time operation of a large power grid. The key features, performance and innovation of the developed LFODA system are introduced below:

(a) It takes massive PMU measurement channels as inputs in both offline and online modes and reports detected oscillation events along with analyzed results as outputs.
(b) It implements a novel Extended Kalman Filter based oscillation detection and analysis approach, which has good performance even with low noise-to-signal ratio.
(c) It supports three effective oscillation detection algorithms including the well-known Prony method, HTLS method and the novel EKF method.
(d) It adopts a novel voting schema among multiple algorithms and an efficient computational flowchart to reduce computational cost. In most cases, the computational expense, compared with crosschecking results among multiple oscillation detection methods, can be reduced by 50%. This characteristic makes LFODA system able to handle many channels of PMUs in real-time.
(e) The detected candidate oscillation modes are verified by a time-series filter in order to ensure consistency of estimation results and further reduce the number of false alarms.
(f) For properly handling data noises, LOFDA is equipped with the DBSCAN method to classify oscillation modes and to group their corresponding PMU measurement points.
(g) The LOFDA system has been installed in a large power utility with 176 PMUs and 1000+ channels, and running continuously for more than 6 months. It demonstrates outstanding performance in detecting low-frequency grid oscillations.

The remainder of this paper is organized as follows. In Section II, three effective oscillation detection methods are introduced. In Section III, an ensemble filter that includes an enhanced voting schema and time-series filter to ensure detection and analysis accuracy is presented. In Section IV, an oscillation mode clustering method by DBSCAN is introduced. In Section V, the LFODA system framework, design and software implementation are presented; its outstanding performance is then tested by case studies using both event and ambient data. Finally, in Section VI, conclusions are drawn with future work identified.

## II. EFFECTIVE OSCILLATION DETECTION METHODS

In this section, three effective oscillation detection methods are introduced to detect oscillations using PMU data, including EKF, Prony and HTLS.

### A. EKF Based Approach

In a discrete framework, a system measurement can be expressed as follows:

$$y_m[k] = \sum_{l=1}^{L} A_{l,m} \exp(-\frac{\sigma_l k}{f_s}) \cos(\frac{\omega_l k}{f_s} + \phi_{l,m}) + \varepsilon_m[k] \qquad (1)$$

where $y_m[k]$ is the measurement of the $m^{th}$ PMU at the $k^{th}$ time instant, $T$ is the transpose operator, $M$ is the number of PMUs, $A_{l,m} \in \mathbb{R}$ is the amplitude, $L$ is the number of oscillation modes, $\sigma_l$ is the damping factor, $\omega_l$ is the frequency, $\phi_{l,m}$ is the phase angle, $f_s$ is the sampling rate, and $\varepsilon_m[k]$ is the measurement error.

The system states include signal magnitudes, frequencies, and damping factors, as follows:

$$x_{l,m}[k] = \begin{bmatrix} x_{l,m}^c[k] \\ x_{l,m}^s[k] \end{bmatrix} = \begin{bmatrix} B_{l,m}^c \exp(-\sigma_l k / f_s) \cos(\omega_l k / f_s) \\ B_{l,m}^s \exp(-\sigma_l k / f_s) \sin(\omega_l k / f_s) \end{bmatrix},$$
$$\omega_l[k] = \omega_l,$$
$$\sigma_l[k] = \sigma_l. \qquad (2)$$

where $B_{l,m}^c \triangleq A_{l,m} \cos(\emptyset_{l,m})$ and $B_{l,m}^s \triangleq -A_{l,m} \sin(\emptyset_{l,m})$.

From (1) and (2), it can obtain the observation function as follows:

$$y[k] = Hx[k] + \varepsilon[k] \qquad (3)$$

$$H = \begin{bmatrix} H_1 \\ H_2 \\ \vdots \\ H_M \end{bmatrix} = \begin{bmatrix} 1\,1\cdots & 0\,0\cdots & \cdots & 0\,0\cdots & 0\,0\cdots \\ 0\,0\cdots & 1\,1\cdots & \cdots & 0\,0\cdots & 0\,0\cdots \\ \vdots & \vdots & \vdots & \vdots & \vdots \\ 0\,0\cdots & 0\,0\cdots & \cdots & 1\,1\cdots & 0\,0\cdots \end{bmatrix}. \quad (4)$$

The constructed system is summarized as follows:
$$\begin{aligned} x[k+1] &= f(x[k]) + \varepsilon[k] \\ y[k] &= Hx[k] + \epsilon[k] \end{aligned} \quad (5)$$

An extended Kalman filter can be applied to estimate the system states using Eq. (5). Kalman filter (KF) is a recursive algorithm to estimate the states of a linear dynamic system using models and a series of noisy measurements. The KF predicts the priori states into the future and computes the difference between the prediction and the measurements based on the dynamic model. Then KF updates the posteriori estimation using the optimal Kalman gain and repeats the process. Kalman filter minimizes the mean squared estimation error with white noises. When the dynamic system is nonlinear, the extended Kalman filter can be applied. Around the current estimated state, the EKF approximates a nonlinear system by a first-order linearization and applies the KF to the linearized system to find the optimal Kalman gain. The nonlinear system model and new measurements are used to calculate new state predictions. This process iterates and the state space model is re-linearized around updated state estimates.

The EKF approach can directly estimate oscillation frequencies and damping factors of multiple modes for online implementation. Its accuracy and stability of the estimates are equal or even better than Prony method when signal-to-noise ratio is low, e.g. 20db. However, its performance relies on the choice of initial points. A fast Fourier Transform (FFT) or other similar technology can be employed as a trigger. Feeding the period of data containing potential oscillation modes into EKF can also increase its performance. The detailed introduction of EKF and its performance can be found in our previous work [16,17].

*B. Prony Method*

In most cases, a power system is a high-order nonlinear and dynamic system. But when system is operated at a quasi-steady condition, e.g., with small disturbances like load changes, the system model can be linearized at its equilibrium point with the following differential and algebraic equation (DAE):
$$\begin{cases} \Delta \dot{x} = A\Delta x + b\Delta u \\ \Delta y_i = c_i \Delta x \end{cases}, i = 1,2,...,m \quad (6)$$

where $\Delta x$ is the system state variable vector; $b$ and $c$ are the input and output coefficient vectors; $\Delta u$ is the input vector; $\Delta y_i$ is the output vector. When the sampling frequency of $y_j(t)$ is $\Delta t$, the discrete form is:
$$y(k) = \sum_{i=1}^{n} R_i z_i^k \quad (7)$$

where $z_i = \exp(\lambda_i \Delta t)$ and $\lambda_i = \delta_i + j\varepsilon_i$. And $n$ is the order of the model. Expanding Eq. (7) yields:

$$\begin{bmatrix} y(0) \\ y(1) \\ \cdots \\ y(N-1) \end{bmatrix} = \begin{bmatrix} 1 & 1 & \cdots & 1 \\ z_1 & z_2 & \cdots & z_n \\ \cdots & \cdots & \cdots & \cdots \\ z_1^{N-1} & z_2^{N-1} & \cdots & z_n^{N-1} \end{bmatrix} \quad (8)$$

To satisfy (9), $z_i$ s are necessarily the roots of $n^{th}$ -order polynomial with unknown coefficient $a_i$.
$$z^n - (a_1 z^{n-1} + a_2 z^{n-2} + \cdots + a_n z^0) = 0 \quad (9)$$

$$\begin{bmatrix} y(n) \\ y(n+1) \\ \cdots \\ y(N-1) \end{bmatrix} = \begin{bmatrix} y(n-1) & y(n-2) & \cdots & y(0) \\ y(n) & y(n-1) & \cdots & y(1) \\ \cdots & \cdots & \cdots & \cdots \\ y(N-2) & y(N-3) & \cdots & y(N-n-1) \end{bmatrix} \begin{bmatrix} a_1 \\ a_2 \\ \cdots \\ a_n \end{bmatrix} \quad (10)$$

Left-multiplying $[0, -a_n, -a_{n-1}, \cdots, -a_1, 1, 0, \cdots 0]$ on both sides of Eq. (8) will yield Eq. (10). The main steps in the Prony algorithm can be summarized: (a) calculating parameter $a_i$ using Eq. (10); (b) obtaining the root of Eq. (9) to get $z_i$; and (c) solving Eq. (8) to get the residual $R_i$.

*C. Hankel Total Least-Squares (HTLS) Method*

The principle of another effective algorithm for detecting oscillations, HTLS, is described below [18].
$$H = \begin{bmatrix} x(0) & x(1) & \cdots & x(M-1) \\ x(1) & x(2) & \cdots & x(M) \\ \cdots & \cdots & \cdots & \cdots \\ x(L-1) & x(L) & \cdots & x(N) \end{bmatrix} \quad (11)$$

A Hankel matrix can be built by Eq. (11). $L > N$ and $M = N+1-L$; $N$ presents the number of points. If the data without noise, then $x(k) = \sum_{i=1}^{n} R_i z_i^k$.

The Hankel matrix can be processed by singular value decomposition that is shown in Eq. (12).
$$H = U \Sigma V^H = \begin{bmatrix} \tilde{U} & U_0 \end{bmatrix} \begin{bmatrix} \tilde{\Sigma} & 0 \\ 0 & \Sigma_0 \end{bmatrix} \begin{bmatrix} \tilde{V} \\ V_0 \end{bmatrix}^H \quad (12)$$

where $U$ and $V$ are unitary matrices. $\tilde{\Sigma}$ is a submatrix and contains the first $n$ singular values and $\Sigma_0$ represents a matrix of non-zero singular values since the data contains noise.

In the absence of noise, $H = \tilde{U}\tilde{\Sigma}\tilde{V}^H$.
$$\tilde{U}_\uparrow = S_\downarrow ZQ = \tilde{U}_\downarrow Q^{-1}ZQ = \tilde{U}_\downarrow \tilde{Z} \quad (13)$$

where $\tilde{Z} = Q^{-1}ZQ$, Q is an n-by-n nonsingular matrix and Z has the same eigenvalues, namely $z_1, z_2, ..., z_n$. In the presence of noise, equation (13) does not always hold. Least square is needed to solve for $\tilde{Z}$. After $\tilde{Z}$ is obtained, its pole is the characteristic value of $\tilde{Z}$. The same method can be easily extended to multi-channel algorithms.

III. THE PROPOSED INNOVATIVE ENSEMBLE FILTERING PROCESS IN THE LFODA SYSTEM

In order to meet the real-time computational goal and ensure robust estimation of oscillation modes, an ensemble filtering process is developed and implemented in the LFODA system for real-time grid operation, which contains two key functions: (a) a novel voting schema to reduce the overall computational cost and speed up the progress of oscillation detection; and (b) a time-series filter to reduce the number of false alarms.

*A. A Novel Voting Schema*

As indicated in [10], [11] and [16, 17] all three methods for oscillation detection including Prony, HTLS, and EKFA can be time-consuming, especially when they are used to handle a large number of signals in real time. In order to improve computational efficiency and make the LFODA system operate in real-time, a highly efficient calculation logic is necessary.

In general, to decide by a group of individuals (say, A, B and C), they will first make their own judgment separately but simultaneously. Then, the final decision of this group is made based on the Majority Rule. However, the percentage of data with oscillation event data with respect to all PMU measurements is relatively rare. It is obvious that the "cross-checking" method based on the Majority Rule becomes an extremely computational costly and time-consuming task. To solve this problem, an improved voting schema is developed, shown in Fig. 1. Instead of waiting for decisions from all candidates A, B and C, the proposed voting schema lets A make judgment first. If A denies, the final decision is rejected. In case A approves, B will be the next to decide. If B also approves, then the final decision is "approved" without C's decision. Otherwise, the final group decision is based on C's judgment.

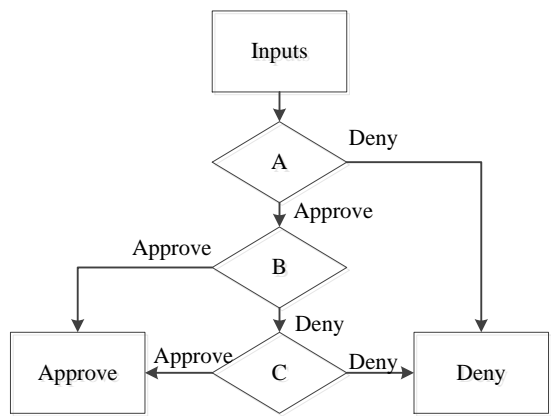

Fig.1. The enhanced voting schema

The LFODA system now supports all three algorithms for oscillation detection. Because most of PMU measurements are ambient data without obvious oscillation modes, Prony method is selected as the first voting member, where the other two methods are used in a standby mode. In this way, the total computational expense can be reduced significantly.

*B. A Time-series Filter to Reduce False Alarms*

For developing an ideal online oscillation detection system, false alarms should be minimized and event alerts should only be triggered when the detected oscillation modes remain consistent. Any method with high sensitivity reporting too many false oscillation events would not be desired by grid operators. Once detected, an oscillation mode of interest should have consistent characteristics during a time window, especially its oscillation frequency. If the frequency changes significantly during two consecutive detection windows, either the previous oscillation mode disappears or is merged into the current mode. Neither case is necessary to trigger the oscillation alarm. In case of improper handling, massive false alarms will be triggered and the system performance is significantly impacted. Therefore, a time-series filter is implemented in the LFODA system to make sure the detected oscillation modes are consistent during a certain period so that false alarms can be effectively reduced without sacrificing accuracy.

By processing massive PMU measurement data containing different levels of noise, it is noticed that a time-series filter can significantly reduce the false alarm by crosschecking detected potential oscillation modes between two continuous detection windows. If the characteristics of a suspected oscillation mode change significantly, the oscillation mode detected in previous detection window should be ignored. Adding this time-series filter can maintain the accuracy of oscillation detection while reducing false alarms.

## IV. AN OSCILLATION MODE CLUSTERING METHOD

To help system operators better understand the detected oscillation events, all events need to be further processed and classified. Firstly, detected oscillations modes are classified into two categories: local and inter-area. For each category, the detected oscillation modes are further classified into different modes in case more than one mode exists. Secondly, PMU measurement channels with oscillation modes are also grouped for displaying and further analysis, such as oscillation source locating, etc.

In reality, the accuracy of PMU measurements can be affected by different factors such as measurement errors from PMU devices and noises from the communication network. The detected oscillation parameters may deviate among different PMU channels even within the same detection period for the same oscillation mode. Similarly, due to measurement noises, the detected oscillation parameters may vary for two continuous detection periods even though the same oscillation mode is being analyzed. Proper grouping of PMU measurements is a challenge indeed. Additionally, the exact number of groups for detected oscillation modes per detection window cannot be known or determined in advance. To address the above challenges, a data mining approach based on the Density-Based Spatial Clustering of Applications with Noise (DBSCAN) method is implemented.

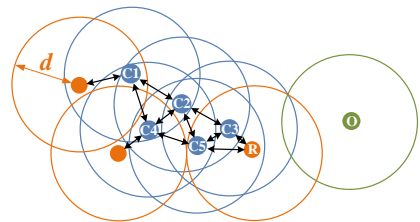

Fig. 2. Schematic diagram of the DBSCAN method with minPts=4

DBSCAN is an unsupervised data mining technique that can classify data points of any dimension into three types: core points, reachable points and outliers [19]. A core point $C$ contains at least *minPts* points (including $C$) within a searching distance *d*. A reachable point *R* exists if there is a path *C1, C2,..., R*, so that all points on the path, except *R*, are core points. Outliers (*O*) are points that are not reachable from any other point. Core points and reachable points can form a cluster without outliers. An example of the DBSCNA method with *minPts*=4 is given in Fig. 2. The detailed implementation of DBSCAN method in LFODA system will be discussed in Section V. A. (3).

## V. Architecture Design, Implementation and Performance of the Developed LFODA System

### A. The Framework of the LFODA System

The LFODA system can consume historical PMU measurements through database or online streams in the form of data packages as inputs, including bus frequencies, voltage magnitudes, voltage phase angles, current magnitudes, current phase angles, active power and reactive power. The outputs of the software system include detected oscillations and their parameters such as oscillation frequency, damping factor, mode shape, etc. The detected oscillations are classified into different modes and the corresponding buses for each mode are also grouped. The main flowchart of the LFODA system is shown in Fig. 3, with three key function modules: (a) real-time data pre-processing module, (b) oscillation detection module and (c) oscillation mode clustering module. All PMU measurement channels are grouped and can be processed in parallel on multi-core workstations to achieve better computational performance.

*1) The real-time measurement pre-processing module*

This module in the LFODA system contains several functions: (a) updating inputs for the LFODA system; (b) removing measurements with errors; and (c) normalization of measurements. Given the fact that LFODA system takes PMU measurements with massive channels in real time, it is important to consider the software system's capability of processing large amount of data, which can be affected by latency from PMU device or communication network [20-22]. As a result, the LFODA system needs to check the availability of updates for each PMU measurement channel in order to avoid inserting historical measurements by mistake.

Since all three oscillation detection methods discussed above can be sensitive to measurement noise, those PMU measurements with obvious errors can adversely affect the performance of oscillation detection. For example, repeated identical historical data for an extended period of time, an example of which is shown in Fig. 4, can lead to a false alarm. It is necessary to detect and remove such PMU measurements with errors.

Normalization of PMU measurements is another useful and important step in the software system, since using normalized values for computation can simplify the calculation of oscillation detection methods.

*2) The oscillation detection module*

The oscillation detection module consists of two functions: (a) oscillation detection for each PMU measurement channel, and (b) false alarm prevention blocks that implement the enhanced voting schema and time-series filter introduced in Section 3. According to [16], Prony is the most sensitive method to detect oscillation modes among these three methods, it is implemented to scan PMU data firstly to avoid missing detection, though the PMU data noises may cause false alarms. HTLS has the similar performance as Prony. In contrast, the EKF approach can provide more accurate results, when the PMU data contains potential oscillation modes even with low signal-to-noise ratio. Besides, the EKF approach can provide better performance when provided with event initial points. As a result, the EKF approach is implemented only to scan the PMU data with potential oscillation modes. The output of this module contains detected oscillation parameters, e.g., oscillation frequency, damping magnitudes, damping factors, phase angles, etc.

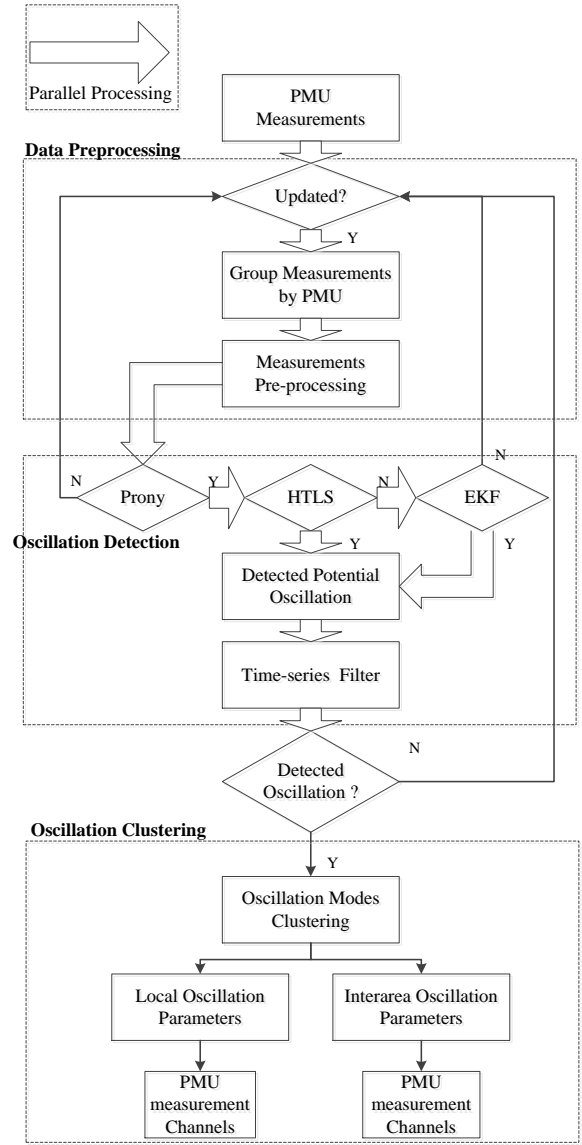

Fig. 3. Main flowchart of the LFODA system

*3) The oscillation mode clustering module*

To enhance the visibility of the oscillation detection results, all detected oscillations are further classified into different types and modes. PMU measurement channels with similar oscillation modes are also classified for further processing and display.

In the LFODA system, oscillation modes are clustered by its frequency. Parameters for the DBSCAN method are set as:

$minPts=1$ and $d=|C|*3\%$. For example, if the frequency of the detected oscillation mode equates to 1 Hz, other oscillation modes with frequencies from 0.97 Hz to 1.03 Hz will be clustered as the same oscillation mode.

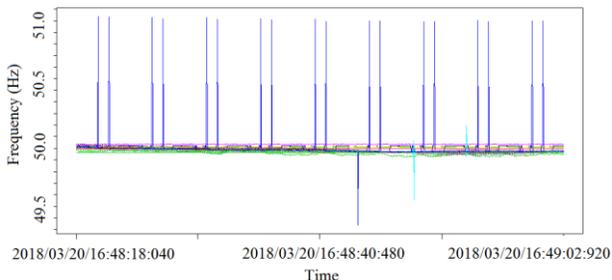

Fig. 4. PMU measurements with periodic repeated historical data

### B. Software Implementation and User Interface of the LFODA System

The core function modules of the LFODA system were primarily developed using C++ programming language (as back end) with the main user interface developed using Javascript (as front end). This software tool was installed in a large power utility in January 2018 and has been running continuously for performance testing since then. This utility owns a total number of 176 PMU devices covering majority of its 220 kV and above transmission network. Currently, more than 1000 measurement channels are being monitored by its control center.

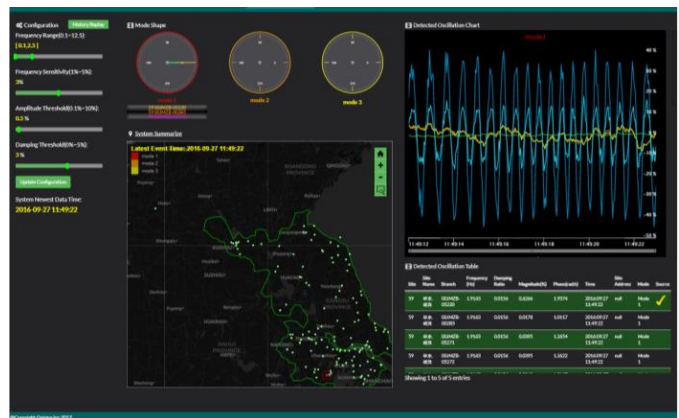

Fig.5. The main user interface of the LFODA system enabled via web browser

The main user interface is shown in Fig. 5, with three function blocks: (1) monitoring software parameter configuration, (2) real-time low-frequency oscillation monitoring and display, and (3) historical event retrieval and playback. Such a design can provide system operators with effective real-time and historical information about grid oscillations, modal analysis, location of oscillation and others. Software users, e.g., grid operators, can manually set the coefficients of the back-end software on the left panel, by dragging the adjustment bar or clicking the yellow area and typing in a target value. Parameters include targeted frequency range, frequency sensitivity, amplitude threshold and damping threshold for low-frequency oscillations.

The top right corner shows oscillating active power curves of selected substations. These curves are clustered according to the oscillation mode of interest. At the bottom right corner, detailed information of the substations with oscillation detected, including names of substations and transmission lines, oscillation frequencies, damping ratios, damping magnitudes, phasor, time, and addresses are listed. The map in the middle shows the oscillation location of the latest oscillation event, where green dots indicate substations without oscillation events and other colored dots represent substations with oscillations. Dots are color coded to differentiate substations with various oscillation modes. The corresponding oscillation mode shape is located above the geographical map. Three oscillation modes can be shown. The display colors are consistent with those substations in the map. In each mode display, the length of each arrow shows the oscillation magnitude and the angle indicates the oscillation phasor.

### C. Case Studies and Findings

The performance of the LFODA system was tested in a few case studies. First, simulated PMU data were used to test the oscillation detection function, including both local and inter-area oscillation modes. In the second study, real-world PMU measurements over an extended period of time were used to test the performance (such as accuracy, speed and robustness) of the developed LFODA system.

*1) Performance on simulated PMU data*

Two test cases from the "Test Cases Library of Power System Sustained Oscillations" [23] are used to test LFODA performance. Both cases were created by running dynamic simulations based on a reduced WECC 179-bus, 29-machine system model. The sampling rate of the simulated PMU signal is 30 Hz and the length of the signal is 5 seconds in each case. In Case I, there is one local oscillation mode with frequency of 1.4 Hz. In Case II, an inter-area oscillation mode with frequency at 0.37 Hz is included. The time-domain signals of the 179 buses monitored by PMUs for both cases are shown in Fig. 6.

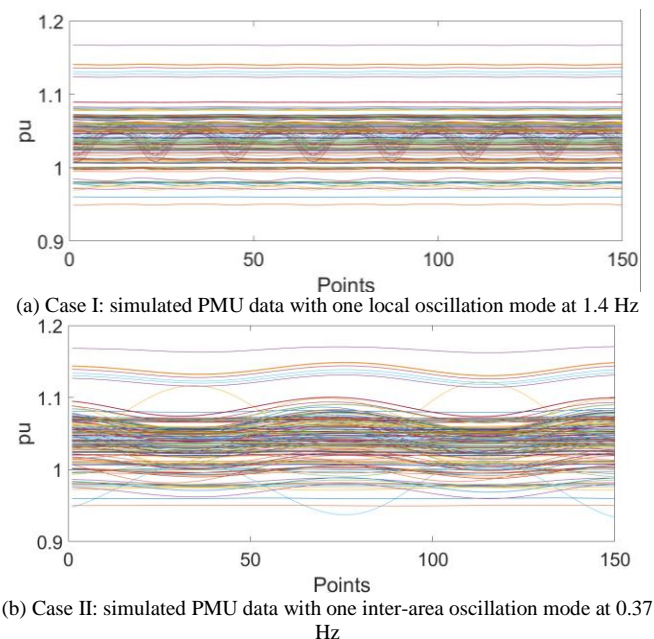

(a) Case I: simulated PMU data with one local oscillation mode at 1.4 Hz

(b) Case II: simulated PMU data with one inter-area oscillation mode at 0.37 Hz

Fig. 6. Case studies with simulated PMU data

The oscillation detection performance of LFODA for both cases is shown in Table 1 and Table 2, respectively. In Case I, there are 13 buses in total that are detected with the exact local oscillation mode. In Case II, five generator buses are reported with an inter-area oscillation mode. Both test results are very similar to the true values published in [23]. Fig. 7 shows the mode shapes of the detected oscillations for both cases, which present the relationship of oscillation's magnitude and phase angle at the related buses. The computational time of the LFODA system for a single time step (1 second) is less than 700 ms, on a workstation (A840-G10) with 4 cores at 2.8 GHz and 256 GB RAM. Both case studies using simulated data proved that the LFODA system can accurately detect and analyze oscillation events in real time.

TABLE 1. OSCILLATION DETECTION RESULTS FOR CASE I

| Type | Frequency (Hz) | Bus ID |
|---|---|---|
| Local Oscillation | 1.4010 | 4; 10; 11; 14; 22; 40; 53; 68; 87; 106; 131; 147; 165 |

TABLE 2. OSCILLATION DETECTION RESULTS FOR CASE II

| Type | Frequency (Hz) | Bus ID |
|---|---|---|
| Inter-area Oscillation | 0.3703 | 6; 9; 13; 47; 79 |

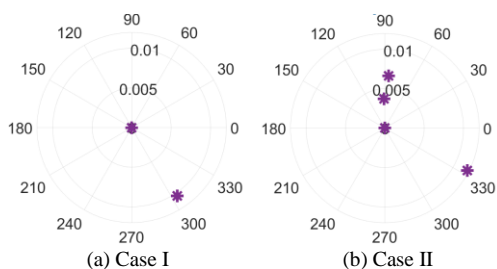

(a) Case I     (b) Case II
Fig. 7. Mode shapes of the detected oscillations

*2) Performance on real-world PMU measurements*

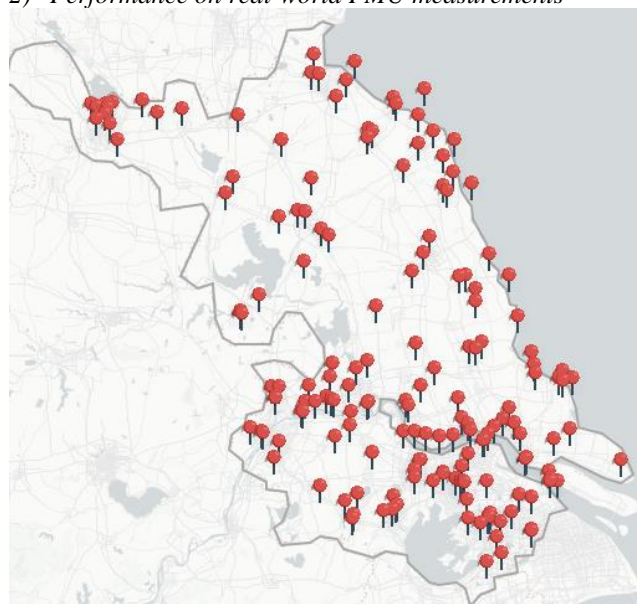

Fig. 8. PMU coverage of the power utility

Three additional cases (Case III, Case IV and Case V) using actual PMU measurements are used to test the accuracy, speeand robustness of the LFODA system. A series of PMU data were collected from this large power utility, with a service area larger than 39.6k sq. miles. The sampling rate of all the PMU measurements is 25 Hz (one sample per 40 ms). The coverage of PMUs on this power utility is shown in Fig. 8.

a) <u>Case III</u>: one historical oscillation event testing

In Case III, PMU measurements containing a historical oscillation event are extracted and used as the inputs to the LFODA system. These measurements contain one local oscillation mode. The original PMU measurements and outputs of the LFODA system are shown in Fig. 9 and Table 3, respectively. It can be observed that one local oscillation can be effectively detected by the LFODA software.

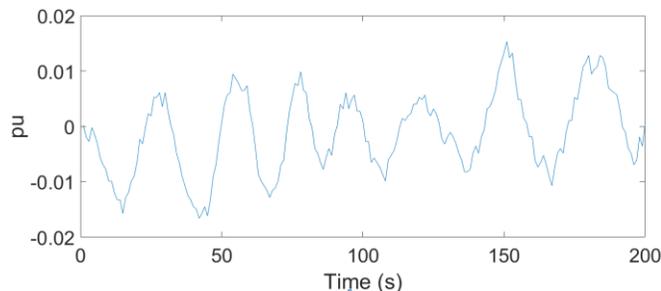

Fig. 9. The actual PMU measurement with one local oscillation mode

TABLE 3. OSCILLATION DETECTION RESULTS FOR CASE III

| Type | Frequency (Hz) | Line ID |
|---|---|---|
| Inter-area Oscillation | 0.9878 | 000-ZPb |

b) <u>Case IV</u>: two years of historical PMU data testing

To further evaluate the performance of the LFODA system, two years (2015 and 2016) of historical real-world PMU measurements are tested. Firstly, the time-series filter in the LFODA system is turned off. More than 20 oscillation events are reported for the first couple months. By analyzing these reported events, most of them last less than one second and are defined as false alarms. By turning on the time-series filter, there are only ten oscillation events that are detected for the two-year period of historical data. Time-series filter can effectively reduce the number of false alarms.

The testing results from the LFODA system are shown in Table 4. Two events have more than one oscillation modes. By replaying and analyzing the historical event data, all the detected oscillation events are confirmed. It can be concluded that the performance of the LFODA system is satisfactory.

TABLE 4. OSCILLATION DETECTION RESULTS FOR 2015 AND 2016

| Time | Type | Frequency(Hz) | PMU_ID |
|---|---|---|---|
| 04-XX-2015 06:04:26 | Local Oscillation | 1.1461 | HRT0DTC1V |
| 04-XX-2015 18:45:40 | Inter-area Oscillation | 0.66 | SNR00481M |
| 06-XX-2015 09:06:48 | Local Oscillation | 1.33 | SNR00461B |
| 01-XX-2016 13:16:09 | Inter-area Oscillation | 0.9521 | BSF01001D |
| 01-XX-2016 03:30:25 | Local Oscillation | 1.0754 | BSF01411FQ |
| 01-XX-2016 03:45:19 | Inter-area Oscillation Local Oscillation | 0.7931, 2.2955, 2.4872 | BSF01411FQ |

| | Local Oscillation | | |
|---|---|---|---|
| 02-XX-2016 21:14:45 | Local Oscillation | 1.2104 | BSF01311BF |
| 06-XX-2016 20:21:27 | Inter-area Oscillation | 0.232 | BSF00051CD |
| 07-XX-2016 18:41:00 | Inter-area Oscillation | 0.235 | BSF00051CD |
| 09-XX-2016 07:32:23 | Inter-area Oscillation Local Oscillation Local Oscillation | 0.3929, 1.1711, 1.5738 | BSFJS151DD |

b) <u>Case V</u>: ambient data testing using 144 PMUs

In Case V, by detecting and removing PMU measurements with errors, measurements from 144 PMUs that monitor more than 800 channels are collected as inputs. The length of these measurements for testing is one minute. A portion of these PMU measurements is shown in Fig. 10. The computation time window of the LFODA system is set to one second. For each window of all the measurement channels being studied, the longest processing time takes less than 800 ms, which is faster than real time.

In this case, the power grid of this utility is under normal operating conditions. All PMU measurements consumed by the LFODA system are considered as ambient data. As expected, no oscillation alert is reported by the LFODA system.

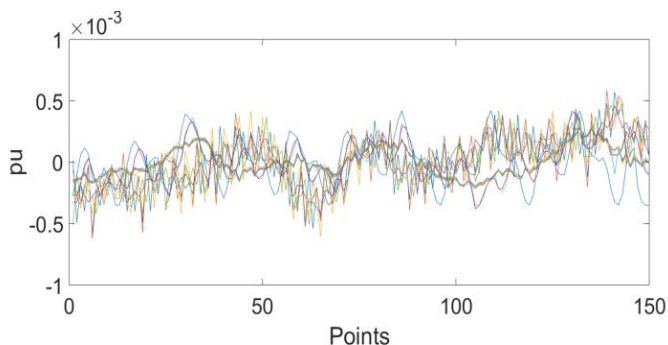

Fig. 10. A portion of PMU measurements with 800+ channels

Fig. 11 compares the total number of calculations for Prony, HTLS and EKFA in the LFODA system and the "crosscheck" method analyzing measurements by all oscillation detection methods. In the LFODA system, Prony is implemented first, followed by the HTLS and KEFA algorithm successively. In the "crosscheck" method, all three oscillation detection algorithms need to be processed for each time step.

For each PMU channel, the Prony algorithm performs one oscillation detection test in both the LFODA system and "crosscheck" method per time step. The total numbers of calculation of Prony are the same for both methods, which is equal to 8,640 calculated using Eq. (22).

$$N_{total} = N_{PMU} \times (\frac{T}{t}) \qquad (22)$$

where $N_{PMU}$ is the total number of PMU, $T$ is the length of the measurement (in second), and $t$ is the time step. Differently, the computational expense for HTLS is significantly reduced in the LFODA system, since the HTLS algorithm only analyzes measurements with potential oscillation modes reported by Prony. Similarly, in the LFODA system, EKFA algorithm only analyzes measurements which Prony and HTLS have different judgments. Thus, the computational cost can be effectively reduced by more than 50%.

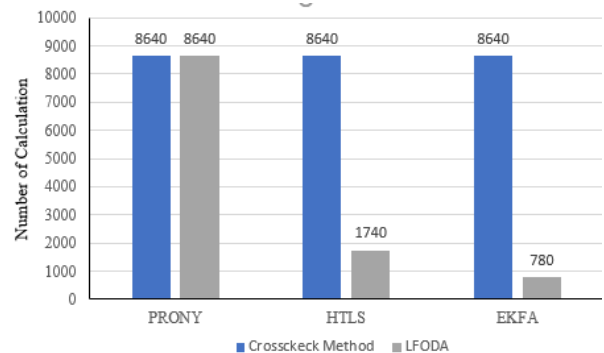

Fig. 11. Comparison of the total calculation cost between the "Crosscheck" method and LFODA

## VI. CONCLUSION AND FUTURE WORK

This paper presents an online LFODA system implemented with a novel EKF approach that has good performance when signal-to-noise ratio is low. It also has an innovative ensemble filtering process to enhance computational speed and oscillation detection accuracy. During oscillations, the LFODA system can detect and analyze oscillations rapidly and accurately. DBSCAN method is adopted to classify detected oscillations into different modes. For each mode, the correlated measurement points are further grouped together for processing and visualization. At normal operating conditions, the LFODA system can effectively reduce false alarms by executing the proposed ensemble filtering process. From the four case studies, the LFODA system has high calculation efficiency by applying the enhanced voting schema and in some cases, the computational cost can be reduced by more than 50% compared to the "crosscheck" method.

In future work, the PMU measurement pre-processing module can be further improved by applying various high-level performance bad data detection and filtering algorithms. Additionally, the oscillation source locating function module can be merged into the LFODA system.

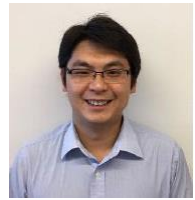

**Desong Bian** (S'12 - M'18) received his B.S. degree from Department of Electrical and Computer Engineering, Tongji University, Shanghai, China in 2007, M.S. degree from Department of Electrical and Computer Engineering, University of Florida, Gainesville, FL, USA in 2011, and Ph.D. from the School of Electrical and Computer Engineering, Virginia Tech, Arlington, VA, USA in 2016. He is currently an Engineer with GEIRI North America, San Jose, CA, USA. His research interests include PMU related applications, demand response, communication network for smart grid, etc.

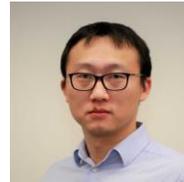

**Zhe Yu** (S'12-M'19) received his B.E. degree from Tsinghua University, Beijing, China in 2009, M.S. degree from Carnegie Mellon University, Pittsburgh, PA, USA in 2010, and Ph.D. degree from Cornell University, Ithaca, NY, USA in 2016 in electrical engineering, respectively. He joined Global Energy Interconnection Research Institute North America (GEIRI North America) in 2017. His current research interests focus on power system and smart grid, demand response, dynamic programming, data processing, and optimization.

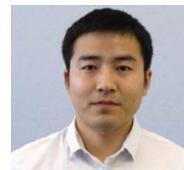

**Di Shi** (M'12-SM'17) received the B. S. degree in electrical engineering from Xi'an Jiaotong University, Xi'an, China, in 2007, and M.S. and Ph.D. degrees in electrical engineering from Arizona State University, Tempe, AZ, USA, in 2009 and 2012, respectively. He currently leads the PMU & System Analytics Group at GEIRI North America, San Jose, CA, USA. His research interests include WAMS, Energy storage systems, and renewable integration. He is an Editor of IEEE Transactions on Smart Grid.

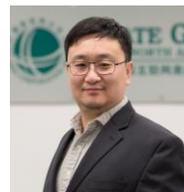

**Ruisheng Diao** ((M'09–SM'15) obtained his Ph.D. degree in EE from Arizona State University, Tempe, AZ, in 2009. Serving as a project manager, PI/co-PI and key technical contributor, Dr. Diao has been managing and supporting a portfolio of research projects in the area of power system modeling, dynamic simulation, online security assessment and control, dynamic state estimation, integration of renewable energy, and HPC implementation in power grid studies. He is now with GEIRINA as deputy department head, PMU&System Analytics, in charge of several R&D projects on power grid high-fidelity simulation tools and developing new AI methods for grid operations.

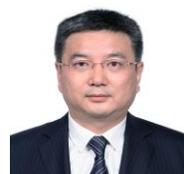

**Zhiwei Wang** (M'16-SM'18) received the B.S. and M.S. degrees in electrical engineering from Southeast University, Nanjing, China, in 1988 and 1991, respectively. He is President of GEIRI North America, San Jose, CA, USA. Prior to this assignment, he served as President of State Grid US Representative Office, New York City, from 2013 to 2015, and President of State Grid Wuxi Electric Power Supply Company from 2012-2013. His research interests include power system operation and control, relay protection, power system planning, and WAMS.